\documentclass{article}
\usepackage{epsf,spconf}
\usepackage{amsmath,amssymb,amsthm,epsfig}

\title {Blind Identification
of Distributed Antenna Systems with Multiple Carrier Frequency
Offsets}

\name {Yuanning Yu, Athina P. Petropulu and H. Vincent Poor$^+$
\thanks{This work was supported by the U. S. National Science Foundation
under Grants
ANI-03-38807, CNS-06-25637 and CNS-04-35052.}}
\address {
Electrical \& Computer Engineering Department,
 Drexel University\\ $^+$School of Engineering and
Applied Science, Princeton University
 }

\begin{document}

\def\bW{\mbox{\boldmath $W$}}

\newcommand{\beq}{\begin{equation}}
\newcommand{\eeq}{\end{equation}}
\newcommand{\beqn}{\begin{eqnarray}}
\newcommand{\eeqn}{\end{eqnarray}}
\newcommand{\om}{\omega}

\sloppy

\ninept

\maketitle

\begin{abstract}

In spatially distributed multiuser antenna systems, the received
signal contains multiple carrier-frequency offsets (CFOs) arising
from mismatch between the oscillators of transmitters and receivers.
This results in a time-varying rotation of the data constellation,
which needs to be compensated at the receiver before symbol
recovery. In this paper, a new approach for blind CFO estimation and
symbol recovery is proposed. The received base-band signal is
over-sampled, and its polyphase components are used to formulate a
virtual Multiple-Input Multiple-Output (MIMO) problem. By applying
blind MIMO system estimation techniques, the system response can be
estimated and decoupled versions of the user symbols can be
recovered, each one of which contains a distinct CFO. By applying a
decision feedback Phase Lock Loop (PLL), the CFO can be mitigated
and the transmitted symbols can be recovered. The estimated MIMO
system response provides information about the CFOs that can be used
to initialize the PLL, speed up its convergence, and avoid
ambiguities usually linked with PLL.

\emph{keywords}-\textbf{Multi-user Systems, Distributed Antenna
Systems, Carrier Frequency Offset, Blind MIMO System Identification}

\end{abstract}

\section{Background}

In both wireless and wireline communication systems, received
signals are often corrupted by carrier-frequency offsets (CFOs), due
to Doppler shift and/or local oscillator drift. The CFO causes a
time-varying rotation of the data symbols, and thus before symbol
recovery, it must be estimated and accurately compensated for by the
receiver. The CFO can be estimated via the use of pilots symbols;
however,
 even a small error in this estimation tends to cause large data
recovery errors. This necessitates transmission of pilot symbols
rather often. In single user systems, or in multiuser systems where
the transmitters are physically connected to the same oscillator,
there is only one CFO that needs to be estimated. This is typically
done via a decision feedback Phase Lock Loop (PLL) at the receiver.
The PLL is a closed-loop feedback control system that can adaptively
track both frequency and phase offsets between the equalized signals
and the reference constellation. However, depending on the
constellation used during transmission, the PLL can have an $M$-fold
symmetric ambiguity, and thus it has limited CFO acquisition range;
e.g., $|f_k|<1/8$ for 4QAM signals. Moreover, the PLL require a long
convergence time. To solve these problems, several methods have been
proposed \cite{Ciblat}, \cite{Ghoto}, \cite{Gini}, \cite{Scott}
\cite{Wang} that allow for blind estimation of the CFO and symbols
using the second-order cyclo-stationary statistics of the
over-sampled received signal. Blind CFO estimation has also been
studied in the context of orthogonal frequency-division multiplexing
(OFDM) systems, where the CFO destroys the orthogonality between the
carriers (see \cite{Koivunen} and the references therein).

In a spatially distributed multiuser antenna system where data are
transmitted simultaneously from multiple antennas, the received
signal contains multiple CFOs, one for each transmit antenna. A PLL
does not work in this case as there is no single frequency to lock
onto. The literature on estimation of multiple CFOs is rather sparse.
In \cite{Frank}, multiple CFOs were estimated by using pilot symbols
that were uncorrelated among the different users. To account for
multiple offsets, \cite{Veronesi} proposed that multiple nodes
transmit the same copy of the data with an artificial delay at each
node. The resulting system was modeled as a convolutive
single-input/single-output (SISO) system with time-varying system
response caused by the multiple CFOs. A minimum mean-square error
(MMSE) decision feedback equalizer was used to track and equalize
the channel and to recover the input data. Training symbols were
required in order to obtain a channel estimate, which was used to
initialize the equalizer.

In this paper, a new approach to blind CFO estimation and symbol
recovery is proposed. The received base-band signal is over-sampled,
and its polyphase components are used to formulate a virtual MIMO
problem. By applying blind MIMO system estimation techniques, the
system response can be estimated, and decoupled versions of the user
symbols can be recovered, each one of which contains a distinct CFO.
By applying a PLL, the CFO can be mitigated and the transmitted
symbols can be recovered. The estimated MIMO system response
provides information about the CFOs that can be used to initialize
the PLL, speed up its convergence, and avoid ambiguities usually
linked with PLLs.

\section{System Model}

We consider a distributed antenna system, where $K$ users transmit
simultaneously to a base station. Narrow-band transmission is
assumed here, where the channel between any user and the base
station is frequency non-selective. In addition, quasi-static
fading is assumed, i.e., the channel gains remain fixed during the
packet length. The continuous-time base-band received signal
$y(t)$ can be expressed as
\begin{equation}\label{cont-system}
 y(t)=\sum_{k=1}^K a_{k} x_k(t-\tau_k) e^{j2\pi F_k t} + w(t) \ ,
\end{equation}
where $a_k$ represents the effect of channel fading between the
$k-$th user and the base station and also phase offset; $\tau_k$ is
the delay associated with the path between the $k-$th user and the
base station; $F_k$ is the frequency offset of the $k-$th user and
$w(t)$ represents noise; $x_k(t)$ denotes the transmitted signal of
user $k$:
\begin{equation}\label{input}
x_k(t)=\sum_i s_k(i)p(t-iT_s) \ ,
\end{equation}
where $s_k(i)$ is the $i-$th symbol of user $k$; $T_s$ is the
symbol period; and $p(t)$ is a pulse function with support
$[0,T_s]$.

Our objective is to obtain an estimate of
${\mathbf{s}}(i)=[s_1(i),...,s_K(i)]^T$ in the form
\begin{equation}\label{20}
{\hat {\bf s}} (i)={\bf {\hat \Lambda} } {\bf P}^T{\bf s}(i) \ ,
\end{equation}
 where $\mathbf{P}$ is a column permutation
matrix and ${\hat {\bf \Lambda}}$ a constant diagonal matrix. These are
considered to be trivial ambiguities, and are typical in any blind
problem.

\section{Formation of the MIMO Problem}

The received signal $y(t)$ is sampled at rate $1/T=P/T_s$, where the
over-sampling factor $P\ge K$ is an integer. In order to guarantee
that all the users' pulses overlap at the sampling times, the over-sampling
period should satisfy: $T_s/P \ge \tau_k, k=1,...K$. Or, in
other words, the over-sampling factor $P$ is upper bounded by
$T_s/\min\{\tau_1,...,\tau_K\}$.

Let $t=iT_s + m T, \ m=1 , \ldots , ,P ,$ denote the sampling times. The
over-sampled received signal can be expressed as {
\hspace{-1.5in}
\begin{eqnarray}
&&y_m(i)=y(iT_s+mT) \nonumber \\
&=& \sum_{k=1}^K a_k e^{j 2 \pi f_k
(i+\frac{m}{P})}x_k((i+\frac{m}{P})T_s-\tau_k) +w((i + \frac{m}{P}
)T_s)
 \nonumber \\
&=& \sum_{k=1}^K a_k e^{j 2 \pi
f_k(i+\frac{m}{P})}s_k(i)p(\frac{m}{P}T_s-\tau_k) +w(iT_s +
\frac{m}{P} T_s)\nonumber \\
&=& \sum_{k=1}^K a_{m,k} (s_k(i) e^{j 2 \pi f_k i}) +w(i +
\frac{m}{P}), \ m = 1, \ldots , P \ , \label{over-sampled}
\end{eqnarray}}
where $f_k=F_k T_s$ is the normalized frequency offset between the
$k-$th user and the base station, and the typical element of the virtual
MIMO channel matrix ${\mathbf A}$ is given by
\begin{equation}\label{virtual-channel}
a_{m,k}=a_k e^{j 2\pi \frac{f_k}{P}m} p\left(\frac{m}{P}T_s-\tau_k\right) \ .
\end{equation}

Define the following: ${\mathbf y}(i) \buildrel \triangle \over
=[y_{1}(i),...,y_{P}(i)]^T$; ${\mathbf A}=\{a_{m,k}\}$, a tall
matrix of dimension $P \times K $; ${\mathbf {\tilde s}}(i)
\buildrel \triangle \over=[s_1(i)e^{j 2 \pi f_1 i},...,s_K(i)e^{j
2 \pi f_K i}]^T$; and ${\mathbf w}(i)\buildrel \triangle
\over=[w(i+\frac{1}{P}),...,w(i+\frac{P}{P})]^T$.
Then, (\ref{over-sampled}) can be written in matrix form as
\begin{equation} \label{matrixform}
{\mathbf y}(i)={\mathbf A {\mathbf {\tilde s}}}(i) +{\mathbf w}(i) \ .
\end{equation}

We could use the training based method of \cite{Frank} to solve the
MIMO system (\ref{over-sampled}). That method assumes that the pilot symbols
of different users are uncorrelated. The CFOs are obtained by
searching for the location of a peak in the cross-correlation
between the Discrete-Time Fourier Transform (DTFT) of a pilot
sequence and that of the received signal.

In the following we show how to estimate CFOs and recover the
transmitted signals in a bind fashion, i.e., without the need for
pilot symbols. The advantage of a blind approach is bandwidth
efficiency since no bandwidth is wasted transmitting pilot symbols.

\section{Blind channel estimation and compensation of the CFOs}

Let us make the following assumptions.
\begin{itemize}
\item
$\mathbf{A1})$ For each $m=1 , \ldots , P$, $w_{m}(.)$ is a zero-mean
Gaussian stationary random processes with variance $\sigma^{2}_{w}$,
and is independent of the channel inputs.

\item
$\mathbf{A2})$ For each $k$, the sequence $s_{k}(i)$ is a zero mean
with independent and identically distributed (i.i.d.) elements
having nonzero kurtosis; i.e., $\gamma_{s_k}^4= \mbox{Cum}[s_{k}(i),
s^{*}_{k}(i), s_{k}(i), s^{*}_{k}(i)]\ne 0.$ The sequences $s_k$'s
are also mutually independent.

\item $\mathbf{A3})$ The over-sampling factor $P$ is no less than $K$.
\end{itemize}
Under assumption (A2), it is easy to verify that the rotated input
signals ${\tilde s}_k(.)$ are also zero mean and i.i.d with nonzero kurtosis.
Also, the ${\tilde s}_k(i)$'s are
mutually independent for different $k$'s. Assumption (A3) guarantees
that the virtual MIMO channel matrix $\mathbf A$ in
(\ref{matrixform}) has full rank with probability one. If the delays
of users are randomly distributed in the interval $[0,T_s/P)$, then
each row of the channel matrix can be viewed as having been drawn randomly from
a continuous distribution so that the channel matrix has full rank
with probability one.

One can apply any blind source separation algorithm (e.g.,
\cite{PARAFAC},\cite{JADE} or \cite{ICA} ) to obtain

\begin{equation}\label{parafacyield}
 {\hat{\bf A}} \buildrel \triangle \over = {\bf A}{\mathbf
P}{\mathbf\Lambda } \ .
 \end{equation}

Subsequently, using a least-squares equalizer we can obtain an estimate
of the de-coupled signals ${\mathbf {\tilde s}}(i)$, within
permutation and scalar ambiguities as
\begin{equation}\label{channelest}
{\bf {\hat {\tilde s}}}(i)= {({\bf \hat{A}}^H \bf \hat{A})^{-1}{\bf
\hat{A}}^H} {\bf y}(i) = e^{jArg \{ {-\bf \Lambda} \}}{\bf |\Lambda|
}^{-1} {\bf P}^T{\bf \tilde s}(i) \ .
\end{equation}
 Without loss of generality we can
assume that the transmitted signal has unit power. Then, on denoting by
$\theta_k$ the $k-$th diagonal element of $Arg\{ {\bf
\Lambda}\}$, the $j-$th separated input signal can be expressed as
\begin{equation}\label{decouple}
{\hat {\tilde s}}_k(i)=s_k(i)e^{j(-\theta_k+ 2 \pi f_k i)} \ .
\end{equation}

In order to recover the transmitted signals, we still need to
mitigate the effect of CFO in each decoupled signal. This can be
done via a PLL. By using the decoupled signals as inputs, and the
constellation used in transmission as a reference, the PLL can
effectively mitigate the CFO by minimizing the feedback error, which
is calculated based on the distance of the recovered signal and the
closest valid constellation point. However, depending on the
constellation used in transmission, there is a four-fold symmetric
ambiguity for MQAM signals, or $M$-fold symmetric ambiguity for MPSK
signals. For example, for 4QAM signals, and an initial CFO value of
$f_k = 0$, the effective tracking range for $f_k$ is $|f_k|< 1/8$.
Moreover, depending on the value of the CFO, the PLL generally needs
a long convergence time, during which the input signals are not
correctly recovered.

Next we will show that by exploring the structure of the virtual
channel matrix, we can obtain an estimate of the CFOs, which can then be
used to initialize the PLL. By doing this, we can prevent the
symmetric ambiguity problem and enlarge the effective tracking range
of the PLL from $|f_k|< 1/8$ to $|f_k|< 1/2$. Also, the
convergence time of the PLL can be greatly reduced.

By taking the phase of the estimated channel matrix $ {\hat {\bf
A}}$, we obtain
\begin{equation} \label{estCFO}
\Psi={\rm Arg}\left\{ {\mathbf {\hat A}} \right\} = \left(%
\begin{array}{ccc}
 \frac{2\pi f_1}{P} + \phi_1 & \ldots & \frac{2\pi f_K}{P} + \phi_K \\
 \vdots & \ddots & \vdots \\
 \frac{2\pi f_1 P}{P} + \phi_1 & \ldots & \frac{2\pi f_K P}{P} + \phi_K \\
\end{array}%
\right) {\mathbf P}
\end{equation}
where $\phi_k={\rm Arg}\{a_k\}+\theta_k$, which accounts for both the
phase of $a_k$ and the estimated phase ambiguity in
(\ref{decouple}).

By applying linear fitting on the $j-$th column of $\Psi$ we obtain
the least squares estimate of $f_j$ as
\begin{equation}\label{estCFO1}
{\hat f}_j=\frac{P}{2\pi}\frac{P (\sum_{p=1}^{P} p
\Psi_{p,j})-(\sum_{p=1}^P p)
(\sum_{p=1}^{P}\Psi_{p,j})}{P(\sum_{p=1}^{P}p^2)- (\sum_{p=1}^{P}
p)^2} \ .
\end{equation}
We can write ${\hat f}_j = f_j+ \epsilon_j$ where $\epsilon_j$
represents estimation error.

On noting that the de-coupled signals ${\hat {\tilde s}}_j(i)$ in
(\ref{decouple}) are shuffled in the same manner as the estimated
CFOs in (\ref{estCFO1}), we can use the estimated CFOs to compensate
for the effects of CFO in the decoupled signals (\ref{decouple}) and
thereby obtain estimates of the input signals as
\begin{equation}\label{recovered}
{\hat {\bf s}}(i)=e^{jArg \{ {-\bf \Lambda} \}} {\bf P}^T{\bf s}(i)
\ .
\end{equation}

Due to errors in the channel estimates, we can only compensate for
most, but not all, of the CFO effects in (\ref{decouple}) and so we
can write
\begin{equation}\label{recovered-error}
{\hat s}_k(i)=s_k(i)e^{j(-\theta_k- 2 \pi \epsilon_k i)} \ .
\end{equation}
By subsequently applying a PLL to ${\hat s}_j(i)$, we can further
mitigate the effects of the residual CFO $\epsilon_k $. For 4QAM
signals, as long as $|\epsilon_k|<1/8$, the residual CFO can be
effectively removed by the PLL. The initial CFO estimator
(\ref{estCFO1}) can prevent the symmetric ambiguity of the PLL, and
can also greatly reduce the convergence time of the PLL. From
(\ref{estCFO}), we can see that the CFO estimator will achieve full
acquisition range for the normalized CFO.

\section{Simulation Results}

In this section, we verify the validity of the proposed method via
simulations, under the following assumptions. The channel
coefficients $a_k, \ k=1, \ldots , K$ are zero-mean Gaussian random
variables. The waveform $p(\cdot)$ is a Hamming window. The delays,
$\tau_k$, $k=1, \ldots , K$ are uniformly distributed in the range
$[0,T_s/P)$. The input signals are 4QAM signals.

The blind source separation algorithm used here is the JADE method,
which was downloaded from
http://www.tsi.enst.fr/\~cardoso/guidesepsou.html.

\begin{figure}[htb4]
\begin{center}
\epsfxsize=3.in \epsfbox{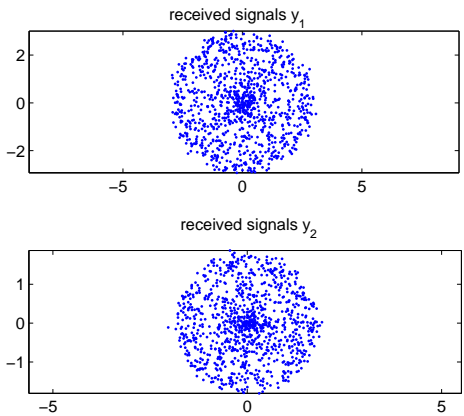} \\
\end{center}
\caption{Received mixing signal ${\bf y}$}\label{received3}
\end{figure}

\begin{figure}[htb4]
\begin{center}
\epsfxsize=3.in \epsfbox{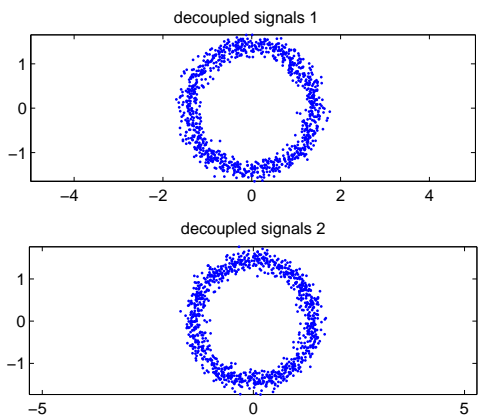} \\
\end{center}
\caption{De-coupled inputs ${\bf {\hat {\tilde
s}}}$}\label{decoupled3}
\end{figure}

\begin{figure}[htb4]
\begin{center}
\epsfxsize=3.in \epsfbox{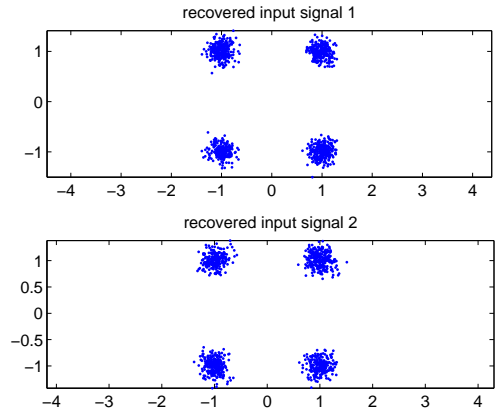} \\
\end{center}
\caption{Recovered input signals ${\bf \hat s}$ with
$P=2$}\label{recovered3}
\end{figure}

First we show results for a two-user systems with $f_1= -0.1552,
f_2= 0.4335$, $a_1=0.3173 - 0.6483i, a_2=0.1625 + 0.5867i$, with
${\rm SNR} =20dB$, and $N=1024$. In Fig. \ref{received3} we show the
polyphase outputs $y_1, y_2$. Due to the mixing and the CFOs no
obvious constellation is visible.
 In Fig. \ref{decoupled3}, we show the de-coupled
signals ${\hat {\tilde s}}_k,$ $k=1,2$ right after JADE. Although
still rotated by the CFOs, two signals ${\tilde s}_k,$ $k=1,2$ are
clearly separated. In Fig. \ref{recovered3}, we show the recovered
input signals ${\hat s}_k,$ $k=1,2$, where we can see that after
compensating for the effect of CFOs the constellations are
recovered.

Next we show estimation results averaged over $300$ independent
channel runs, and $20$ Monte-Carlo runs for each channel. For each
channel case, the coefficients $a_k, k=1,2$ were generated randomly,
and the continuous CFOs where chosen randomly in the range
$[-\frac{1}{2T_s},\frac{1}{2T_s})$. The delays, $\tau_k$,
$k=1,...,K$ where chosen uniformly in the range of $[0,T_s/P)$. The
transmitted signal was 4QAM.

\begin{figure}[htb4]
\begin{center}
\epsfxsize=3.in \epsfbox{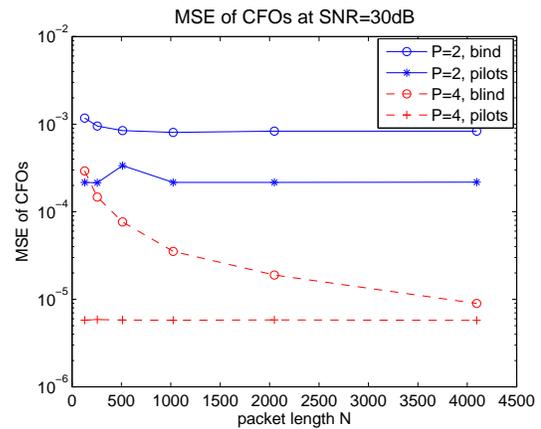} \\
\end{center}
\caption{MSE of CFOs vs $N$ for K=2, with SNR=30dB,
4QAM}\label{CFOMSE1}
\end{figure}

\begin{figure}[htb4]
\begin{center}
\epsfxsize=3.in \epsfbox{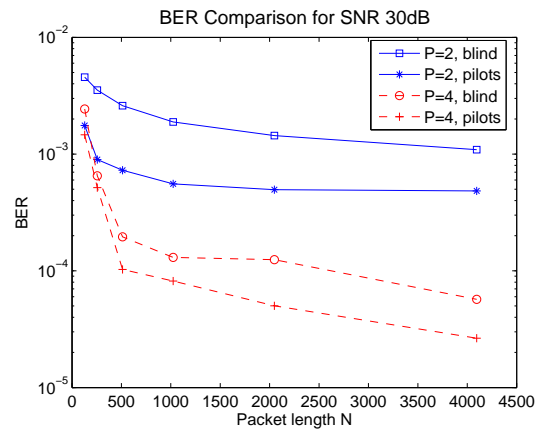} \\
\end{center}
\caption{BER vs $N$ for K=2, with SNR=30dB, 4QAM}\label{BER1}
\end{figure}

The performance of both the pilot-based method and the proposed
method at different data lengths and with SNR set to $30$dB is shown
in Fig. \ref{CFOMSE1}. For the pilot-based method, each user
transmits a pilot sequence of length $32$, and the pilots are random
sequences uncorrelated between different users. Fig. \ref{CFOMSE1}
shows the mean-square error (MSE) for the CFO estimate
(\ref{estCFO1}) for different values of the over-sampling factor
$P$. The MSE is calculated usings $\frac{1}{K}\sum_{k=1}^{K}[({\hat
f}_k-f_k)]^2 = \frac{1}{K} \sum_{k=1}^{K}[({\hat F}_k-F_k)T_s]^2$.
We can see that by increasing $P$ we can improve the estimation
accuracy. Fig. \ref{BER1} shows the Bit Error Rate (BER) for
different values of $P$. For both blind and training methods, the
BER is calculated based on the recovered signals after the PLL. As
expected, the BER performance also improves by increasing $P$. The
proposed method appears to work well even for short data length.

\begin{figure}[htb4]
\begin{center}
\epsfxsize=3.in \epsfbox{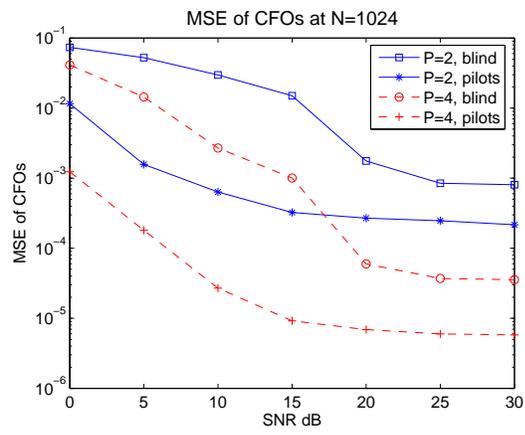} \\
\end{center}
\caption{MSE of CFOs vs SNR for K=2, 4QAM}\label{CFOMSE2}
\end{figure}

\begin{figure}[htb4]
\begin{center}
\epsfxsize=3.in \epsfbox{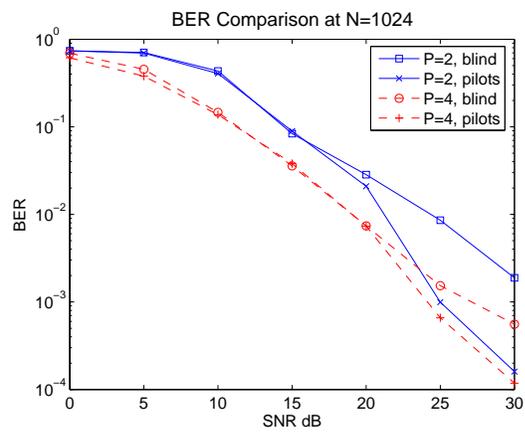} \\
\end{center}
\caption{BER vs SNR for K=2, 4QAM, T=1024}\label{BER2}
\end{figure}

 Next we show the
performance of both methods at various noise levels. We use packet
length $N=1,024$. In Fig. \ref{CFOMSE2} we show the MSE of the blind
CFO estimator (\ref{estCFO1}) as well as that of the training based
method. We can see that by increasing $P$ we improve estimation
accuracy. In Fig. \ref{BER2}, we show the BER performance after the
PLL for both blind and training based methods. We see that the
proposed blind method has almost the same performance as the
training based method for SNR values lower than 20dB, while the
training based method can achieve better BER performance for higher
values of SNR.

\section{Conclusion}

In this paper we have proposed a novel blind approach for
identification of a distributed multiuser antenna system with
multiple CFOs. By over-sampling of the received base-band signal,
the MISO problem is converted into a MIMO one. Blind MIMO system
estimation then yields the system response, and MIMO input recovery
yields the decoupled transmitted signals, each one containing a
CFO. By exploring the structure of the MIMO system response we
obtain a coarse estimate of the CFOs, which is then combined with
a decision feedback PLL to compensate for the
CFOs in the decoupled transmitted signals. The proposed blind
method has full acquisition range for normalized CFOs.


\end{document}